\documentclass[sigconf,nonacm]{acmart}
\usepackage{todonotes}

\usepackage[skins, breakable]{tcolorbox}
\usepackage{xcolor}
\usepackage{upquote}
\usepackage{fontawesome5}
\usepackage[skip=0pt]{caption}
\usepackage{threeparttable}
\usepackage[table,xcdraw]{xcolor}
\usepackage{orcidlink}

\usepackage{hyperref}

\AtBeginDocument{%
  }

\acmConference[ ]{ }

\begin{document}

\title{POLARIS: Is Multi-Agentic Reasoning the Next Wave in Engineering Self-Adaptive Systems? }


\author{Divyansh Pandey}
\authornote{These authors contributed equally to this research.}
\orcid{0009-0002-2915-9624}
\affiliation{%
  \institution{SERC, IIIT - Hyderabad}
  \city{}
  \state{}
  \country{}}
\email{divyansh.pandey@students.iiit.ac.in}

\author{Vyakhya Gupta}
\authornotemark[1]
\orcid{0009-0003-0879-6226}
\affiliation{%
  \institution{SERC, IIIT - Hyderabad}
  \city{}
  \state{}
  \country{}}
\email{vyakhya.gupta@students.iiit.ac.in}

\author{Prakhar Singhal}
\authornotemark[1]
\orcid{0009-0009-7268-5508}
\affiliation{%
  \institution{SERC, IIIT - Hyderabad}
  \city{}
  \state{}
  \country{}}
\email{prakhar.singhal@research.iiit.ac.in}

\author{Karthik Vaidhyanathan}
\affiliation{%
  \institution{SERC, IIIT - Hyderabad}
  \city{}
  \state{}
  \country{}}
\email{karthik.vaidhyanathan@iiit.ac.in}
\orcid{0000-0003-2317-6175}



\begin{abstract}

The growing scale, complexity, interconnectivity, and autonomy of modern software ecosystems introduce unprecedented levels of uncertainty, challenging the foundations of traditional self-adaptation. Existing self-adaptive techniques, often based on rule-driven controllers or isolated learning components, struggle to generalize across novel contexts and coordinate responses across distributed subsystems, leaving them ill-equipped to handle emergent \textit{``unknown unknowns''} in complex, dynamic environments. Recent discussions on \textit{Self Adaptation 2.0} established an equal partnership between AI and adaptive systems, merging learning-driven intelligence with adaptive control to enable predictive, data-driven, and proactive adaptation. Building on this foundation, we introduce a new wave of self-adaptation through \textit{POLARIS} 
a three-layer multi-agentic self-adaptation framework that moves beyond reactive adaptation by combining (1) a low-latency Adapter layer for monitoring and safe execution, (2) a transparent Reasoning layer that generates and verifies plans using tool-aware, explainable agents, and (3) a Meta layer that records experiences and meta-learns improved adaptation policies over time. By using shared knowledge and predictive models, POLARIS can handle uncertainty, learn from its past actions, and evolve its strategies, enabling the engineering of autonomous systems that can anticipate change to ensure resilient and goal-directed behavior under uncertainty.  Preliminary evaluation across two distinct self-adaptive exemplars, SWIM and SWITCH, demonstrates that POLARIS consistently outperforms existing state-of-the-art baselines. With this, we motivate a shift towards \textit{Self-Adaptation 3.0} akin to \textit{Software 3.0}, a new paradigm where systems move beyond merely learning from their environment to reasoning about and evolving their own adaptation. In this vision, adaptation contributes to a self-learning process that enables systems to continuously improve and respond to novel challenges. 

\end{abstract}

\maketitle

\section{Introduction}

Modern software systems are increasingly required to operate autonomously in dynamic and uncertain environments. From large-scale distributed infrastructures to cyber-physical systems, the ability to self-adapt has become essential for maintaining resilience and long-term performance. Traditional self-adaptive systems, typically governed by feedback loops and control-theoretic models, have been successful in managing predictable or modeled uncertainties~\cite{casimiro2024probabilistic,esfahani2011uncertainity2,moreno2018uncertainity1,arcaini2015modeling}. However, the rapid integration of artificial intelligence (AI) into software ecosystems has fundamentally altered this landscape. AI-enabled systems introduce new forms of non-determinism, stemming not only from environmental variability but also from the adaptive and opaque nature of AI components themselves. While existing frameworks have started to incorporate AI-based prediction or decision modules, they largely treat AI as an auxiliary tool within classical control loops. This limits their ability to manage \textit{emergent and evolving uncertainties} such as data drift, stochastic models, or co-evolving agents: scenarios increasingly common in AI-native systems. Moreover, adaptation in such systems must go beyond reactive adjustments; it must \textit{learn from adaptation itself}. Systems should continuously refine their reasoning and decision strategies over time, improving with experience rather than merely responding to change~\cite{kinneer2021}. This need for continual, self-improving adaptation marks a shift toward \textit{AI-native self-adaptation}, where learning and reasoning are integral to the adaptation process.

Building on the vision of \textit{Self-Adaptation~2.0} by Bureš~\cite{bures2021self}, we argue that the next wave of self-adaptive systems~\cite{weyns2018engineering} must treat intelligence as a first-class design principle. In this work, we present \textit{POLARIS}-\textit{Proactive Orchestrated Learning for Agentic and Reasoning-driven Intelligent Systems}, which redefines adaptation as an emergent property of interacting agents. 
POLARIS adopts and builds into a three-layer architecture proposed by Kramer et al.~\cite{3layer}, comprising: (1) an Adapter layer for monitoring and safe execution; (2) a transparent Reasoning layer with tool-aware, explainable agents for planning and verification; and (3) a Meta layer to record experiences and meta-learn improved adaptation policies over time. These three layers combine to give the framework its orchestrated and continuous learning nature.
We evaluate \textit{POLARIS} on two distinct exemplars, \textit{SWIM} \cite{swim} and \textit{SWITCH} \cite{switch}, highlighting its performance and broad applicability across both traditional legacy software systems and modern AI-based ones. On \textit{SWIM}, \textit{POLARIS} achieves a Total Utility value of 5445.48 succeeding the existing baselines. On \textit{SWITCH}, \textit{POLARIS} shows significant efficiency gains over the existing baseline, achieving a 27.3\% lower median response time, 14.9\% lower CPU usage, and an 87.1\% reduction in disruptive switches. These results collectively demonstrate that \textit{POLARIS} not only adapts effectively to diverse system architectures but also delivers measurable improvements in performance and resource efficiency. The complete  replication package is made available .\footnote{\href{https://github.com/sa4s-serc/POLARIS}{https://github.com/sa4s-serc/POLARIS}}


\section{Related Work}
\label{sec:related_work}

The foundations of self-adaptation are built upon structured control loops for autonomic 
behavior, notably the \textit{MAPE-K} model, which relied on a largely centralized and sequential organization~\cite{kephart2003vision,arcaini2015modeling}. To improve modularity and scalability, hierarchical patterns such as the three-layer model~\cite{kramer2007self,weyns2020introduction} distinguished component control, change management, and goal management, while decentralized and multi-loop approaches explored distributed coordination through multi-agent systems (MAS)~\cite{wooldridge2009multiagent}. These frameworks significantly advanced autonomy and resilience, but largely left open how systems can reason about and change their own adaptation mechanisms. Addressing runtime uncertainty has been a central challenge. Techniques like \textit{POISED}~\cite{esfahani2011uncertainity2} and uncertainty-reduction tactics~\cite{moreno2018uncertainity1} improve decision quality under incomplete information, while quantitative approaches that use probabilistic models and model checking~\cite{casimiro2024probabilistic,calinescu2017runtime} reason about adaptation strategies with confidence. Control-theoretic hybrids~\cite{caldas2020hybrid}, predictive frameworks such as \textit{RAINBOW}~\cite{garlan2004rainbow}, and model-based approaches like \textit{CobRA}~\cite{cobrapaper}, with comparative studies validating their effectiveness~\cite{moreno2017predictive}, extend foresight and dependability. Collectively, these contributions form a rigorous foundation for dependable adaptation. However, their primary focus remains on system-level behavior rather than on enabling reasoning or self-evolving strategies. The integration of machine learning (\textit{ML}) marked a transition toward data-driven decision-making. Systematic reviews~\cite{gheibi2021ml} highlight the use of supervised and reinforcement learning for environment modeling and policy optimization. Online reinforcement learning (\textit{RL}) methods have been applied to learn policies at runtime; for example, feature-model-guided exploration speeds up online \textit{RL} in systems with large action spaces or evolving configurations~\cite{kim2009reinforcement, metzger2022}. Other studies combine \textit{ML} with formal analysis~\cite{camara2020quantitative}, and recent work on lifelong adaptation enables strategies to improve with experience~\cite{gheibiLifelong}. \textit{Self-Adaptation 2.0}~\cite{bures2021self} further advocates an AI-native paradigm. Li et al.~\cite{li2024genaistudy} examine this intersection through a survey of Generative AI, with recent works demonstrating its use for tasks like goal-model generation within \textit{MAPE-K}~\cite{nakagawa2023}. Subsequent works like \textit{AWARE}~\cite{chekam2025aware} and \textit{MSE-K}\cite{reimagineLLM} build on this view. Although these works contribute compelling techniques, they employ AI to augment classical control functions without establishing how reasoning becomes an analytical activity that accounts for AI-induced uncertainty. The notion of adaptation as an emergent property of interacting reasoning agents: where cognitive processes themselves become the substrate of adaptation, remains underexplored. In parallel, the rise of \textit{agentic AI} is redefining software architecture. LLMs, seen as autonomous reasoning components~\cite{sapkota2025ai,xi2025rise}, are prompting new architectural patterns~\cite{vaidhyanathan2025software,liu2025agent} and agentic orchestration frameworks~\cite{he2025llm,wang2025agents}. These lines of research point toward systems where prediction, verification, learning, and negotiation are performed by collaborating agents. Building on these trends, POLARIS operationalizes \textit{AI-native self-adaptation} through reasoning-enabled agents that unify predictive foresight and runtime assurance with continual learning within a multi-agentic framework.

\section{Motivation}
Ever since Kephart and Chess proposed the vision of autonomic computing~\cite{kephart2003vision}, the broader goal has been to build systems that can manage themselves with minimal human intervention. The original MAPE-K model laid the foundation for this vision by introducing feedback loops for monitoring, analysis, planning, and execution. But true autonomy requires more than automated control. It demands systems that can reason about their actions, learn from experience, and continually improve their adaptation strategies.

In practice, when faults occur, human experts collaborate, reason through causes, and decide how to respond. Achieving this kind of autonomous reasoning and teamwork in software has been a long-standing problem. Traditional feedback-based approaches remain reactive in nature as they adapt but rarely learn. Bureš et al.~\cite{bures2021self} describe this limitation as the threshold between early self-adaptive systems and what they call Self-Adaptation 2.0, where data-driven and AI techniques make adaptation predictive and proactive. Yet, even Self-Adaptation 2.0 stops short of reasoning and self-improvement. To reach that level, systems must evolve toward Self-Adaptation 3.0, where reasoning, learning, and coordination work together, thereby allowing adaptation itself to adapt. The rise of Agentic AI now makes this vision realistic: a collection of goal-driven LLM-based agents can function like a team of expert engineers, analyzing context, proposing actions, and improving over time. This shift parallels the move from Software 2.0 to Software 3.0 as articulated by Andrej Karpathy ~\footnote{https://www.ycombinator.com/library/MW-andrej-karpathy-software-is-changing-again}. It marks the next leap: systems powered by large language models (LLMs) and reasoning engines that flexibly access, interpret, and orchestrate both symbolic (Software 1.0) and learned (Software 2.0) components. In this era, natural language becomes the new programming interface, and large models act as intelligent engines capable of reasoning, collaboration, and adaptive behavior. Yet, relying solely on a single LLM for reasoning introduces challenges, such as hallucination, lack of context retention, and inconsistent decision quality, necessitating a design that actively guides these models through feedback, improvement, and tool-mediated interaction \cite{hallucination}. As a result, software moves from passive execution toward \textit{reasoning-driven adaptation}, where systems understand, plan, and respond to goals in real time. Incorporating this paradigm into self-adaptive systems calls for a new generation of adaptive intelligence. Such systems must integrate reasoning, continual learning, and distributed coordination to handle the unknowns of open, dynamic environments. These requirements motivate \textit{POLARIS}, which unites proactive reasoning, reflective learning, and coordinated multi-agent intelligence to achieve robust and autonomous adaptation. Taken together, this direction points toward what could be seen as the next stage in the evolution of self-adaptation. POLARIS aims to address the following key challenges:
\\
\textit{CH1: Generalizable, Reasoned Adaptation:} Integrate language-based reasoning, reflective learning, and predictive modeling to anticipate and plan adaptive responses under uncertainty.\\
\textit{CH2: Learning-Driven Evolution:} Capture and reuse experiential knowledge through meta-learning to continuously refine adaptation strategies and decision quality.\\
\textit{CH3: Coordinated Multi-Agent Intelligence:} Enable coherent, goal-aligned collaboration among autonomous agents to achieve distributed yet globally consistent adaptations.

\section{Agentic Preliminaries}
This section establishes the foundational concepts underlying the agentic organization of POLARIS, namely the notions of \textit{agents}, \textit{AI agents}, and \textit{tools}. These definitions provide the conceptual basis for distinguishing between autonomous, supportive, and facilitating entities within the framework.

\smallskip
\noindent \textbf{Agent} - An \textit{agent} is a computational entity that operates autonomously within an environment to pursue defined objectives. According to Wooldridge~\cite{wooldridge2009multiagent}, an agent is ``a computer system capable of autonomous action in an environment to meet its design objectives.'' Agents possess local goals, perceive their operational context, and act upon it to maintain or improve system-level performance. 

\noindent \textbf{AI Agent} - An \textit{AI agent} extends a basic agent by incorporating reasoning, learning, and adaptive decision-making capabilities. Unlike reactive agents that follow fixed rules, AI agents can interpret context, predict outcomes, and refine their strategies over time using AI models as their "cognitive core"~\cite{liu2025agent}.


\noindent \textbf{Tool} - A \textit{tool} refers to a non-autonomous software artifact that supports agents in perception, reasoning, or learning but lacks independent goals or agency. It extends an agent’s capabilities and efficiency\cite{tool}.

\smallskip
\noindent Beyond these, the framework also includes other infrastructural and structural elements that facilitate communication, coordination or execution across the system. These components serve as the bridge enabling agent collaboration and interaction with the managed environment.


\begin{figure*}[h] 
    \centering 
    \includegraphics[width=0.8\linewidth]{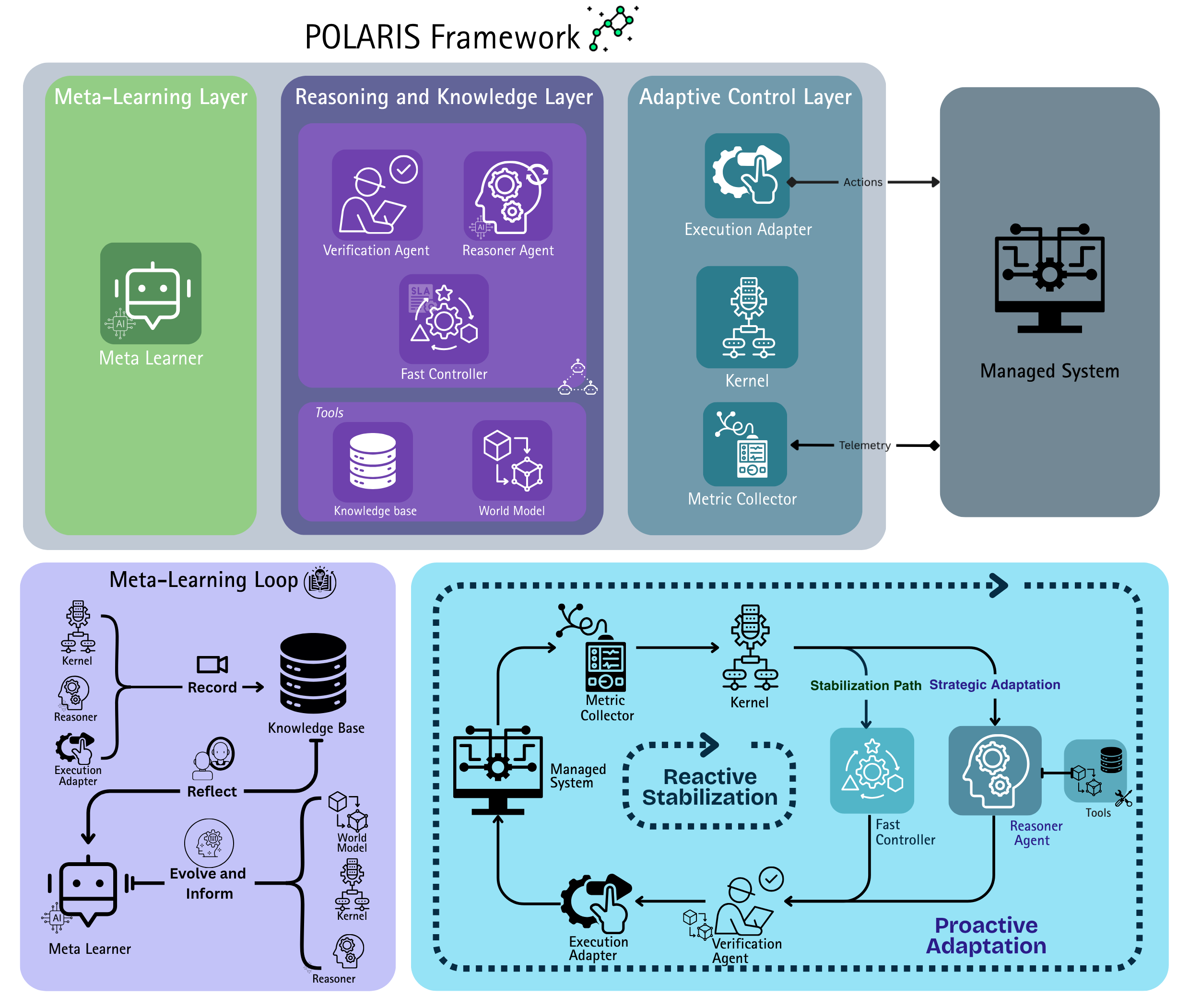} 
    \hspace{-5pt} 
    \caption{POLARIS Framework: The three layers along with adaptation loops visualized}
    \label{methodology} 
\end{figure*}

\section{POLARIS: Overview and Methodology}
\label{sec:methodology}

POLARIS orchestrates self-adaptation through a structured, multi-agent architecture designed for proactive reasoning and continuous improvement. Its design, visualized in Figure~\ref{methodology}, consists of three conceptual layers: \textit{Adaptive Control}, \textit{Reasoning \& Knowledge}, and \textit{Meta-Learning}, that operationalize adaptation across different timescales. This layered approach, inspired by the three-layer reference model~\cite{kramer2007self}, provides a clear separation of concerns, distinguishing immediate system interaction from deliberative reasoning and long-term strategic evolution.

\begin{table}[htbp!]
\centering
\begin{threeparttable}
\caption{Overview of POLARIS Core elements and Responsibilities}
\label{tab:components}
\scriptsize
\renewcommand{\arraystretch}{1.2} 
\setlength{\tabcolsep}{5pt} 
\begin{tabular}{@{}p{0.2\linewidth} p{0.2\linewidth} p{0.54\linewidth}@{}}
\toprule
\textbf{Component} & \textbf{Layer} & \textbf{Primary Responsibility} \\ 
\midrule
\textbf{Kernel ($K$)} & \textit{Adaptive Control} & Central orchestrator; triages system states and dispatches control to the appropriate loop. \\
\textbf{Metric Collector ($MC$)} & \textit{Adaptive Control} & Gathers raw and derived performance telemetry from the managed system. \\
\textbf{Execution Adapter ($EA$)} & \textit{Adaptive Control} & Translates abstract adaptation commands into domain-specific, executable actions. \\ 
\midrule
\textbf{Reasoner ($R$)} & \textit{Reasoning\tnote{*}} & AI-driven agentic system that uses tools for deliberative, evidence-based strategic planning. \\
\textbf{Verifier ($V$)} & \textit{Reasoning} & Acts as a safety backstop, validating proposed plans against operational invariants. \\
\textbf{Fast Controller ($F$)} & \textit{Reasoning} & Executes a predefined, low-latency policy to rapidly stabilize the system during crises. \\
\textbf{Knowledge Base ($KB$)} & \textit{Reasoning (Tool)} & Serves as the system's episodic memory, storing historical experiences and learned patterns. \\
\textbf{World Model ($WM$)} & \textit{Reasoning (Tool)} & Provides a predictive, simulatable model for "what-if", counterfactual reasoning. \\ 
\midrule
\textbf{Meta-Learner ($M$)} & \textit{Meta-Learning\tnote{*}} & Analyzes long-term adaptation history to identify suboptimal patterns and evolve system policies. \\ 
\bottomrule
\end{tabular}
\begin{tablenotes}
\item *Dedicated AI components that require reasoning cores; can instantiate multiple specialized AI agents with local knowledge.
\end{tablenotes}
\end{threeparttable}
\end{table}

\subsection{Operational Flow}
Formally, the POLARIS framework ($\mathcal{F}$) comprises three primary constituent sets that define its structure and capabilities:
\[
\mathcal{A} = \{R, M, V, F\}, \quad \mathcal{T} = \{KB, WM\}, \quad \mathcal{C} = \{K, MC, EA\}
\]
Here, $\mathcal{A}$ represents the set of \textit{active agents} responsible for decision-making: the \textbf{Reasoner} ($R$), \textbf{Meta-Learner} ($M$), \textbf{Verifier} ($V$), and \textbf{Fast Controller} ($F$). $\mathcal{T}$ is the set of \textit{foundational tools} that support deliberation: the \textbf{Knowledge Base} ($KB$) and \textbf{World Model} ($WM$). Finally, $\mathcal{C}$ is the set of \textit{operational components} forming the interface to the managed system: the \textbf{Kernel} ($K$), \textbf{Metric Collector} ($MC$), and \textbf{Execution Adapter} ($EA$).

As depicted in Figure~\ref{methodology}, these components collaborate across three nested feedback loops:
\begin{itemize}
    \item an immediate \textit{ReactiveStabilization} loop
    \item a short-term \textit{ProactiveAdaptation} loop
    \item and a long-term \textit{Meta-Learning} loop 
\end{itemize}

To ground the discussion, we illustrate the architectural flow using the SWIM exemplar~\cite{swim} as a representative case study, where adaptation involves dynamically scaling the servers and adjusting request processing fidelity using \textit{dimmer} control.

\subsection{ Reactive Stabilization}
The innermost loop guarantees immediate, bounded-latency responses to critical system failures, prioritizing stability.

\paragraph*{Trigger and Dispatch.} The process begins with the \textbf{Metric Collector (MC)}, which provides a continuous stream of telemetry. This data flows to the \textbf{Kernel (K)}, the framework's central orchestrator. The Kernel analyzes each metric observation $m$ against a critical threshold $\theta_{critical}$. If a violation is detected (e.g., $m > \theta_{critical}$), the Kernel activates the \textit{Stabilization Path} (illustrated in Fig ~\ref{methodology}), immediately delegating control. For instance, if the average response time $m_{rt}$=900\,\textit{ms} exceeds a $\theta_{critical\_rt}$ of 750\,\textit{ms}, this loop is triggered.

\paragraph*{Reactive Action.} Control is passed to the \textbf{Fast Controller (F)}, a lightweight agent executing a predefined policy. This policy consists of simple, reliable rules that map an observed system state $s$ to a corrective action $a_{react}$:
\[
a_{react} = \Phi(s)
\]
For example, $\Phi$ might map the high response time state to a `set dimmer to 70\%` action. As illustrated in Figure ~\ref{methodology}, this action is routed through a verification stage before execution. It is then passed to the \textbf{Execution Adapter (EA)}, which translates it into a domain-specific command, ensuring the system is rapidly returned to a known-safe state.

\subsection{Proactive Adaptation}
When the system is not in immediate crisis, POLARIS engages its deliberative reasoning capabilities.

\paragraph*{Trigger and Context.} This loop operates as the system's default, continuous adaptive reasoning cycle. The \textbf{Kernel} invokes the \textbf{Reasoner Agent (R)} (using the \textit{Strategic Adaptation} path as shown in Fig ~\ref{methodology}) with a specific reasoning context, derived from observed system metrics and operational state:
\[
P_R = \langle G, \mathcal{I}, \mathcal{T} \rangle
\]
This context defines the task: $G$ specifies optimization goals (e.g., minimize cost), $\mathcal{I}$ represents hard invariants (e.g., $\texttt{max\_servers} \leq 3$), and $\mathcal{T}$ provides tool interfaces. The Reasoner operates autonomously, using telemetry and internal heuristics to detect emerging trends or inefficiencies warranting action.

\paragraph*{Prompt Configuration.}
The Reasoner (which has an LLM agent(s)) operates using a structured \textit{prompt configuration} that encodes system constraints, adaptation thresholds, and reasoning workflow templates. This configuration defines how contextual data, goals, and invariants are mapped into a consistent reasoning process, ensuring interpretable and reproducible decisions. Formally, $\mathcal{P}_{conf} = \langle \mathcal{I}, \mathcal{D}, \mathcal{R}, \mathcal{O} \rangle$, where $\mathcal{I}$ represents fixed invariants, $\mathcal{D}$ defines key operational directives and performance targets, $\mathcal{R}$ outlines reasoning procedures (e.g., Chain of Thought Prompting, Tool call definitions), and $\mathcal{O}$ specifies structured output formats. For instance, $\mathcal{D}$ might define a dangerous utilization level ($\texttt{server\_util\_danger} = 0.90$) or a target response time ($\texttt{response\_time\_goal} = 0.5s$), which the LLM uses to inform its planning. Through $\mathcal{P}_{conf}$, prompts are dynamically instantiated from context $P_R$, maintaining consistent logic across adaptation cycles.
\paragraph*{Example Configuration.}
A skeleton of the prompt for Reasoner
\begin{tcolorbox}[
    enhanced,
    colback=cyan!10!white,                     
    colframe=cyan!50!gray,                     
    boxrule=0.4pt,
    arc=2pt,
    left=5pt,
    right=5pt,
    top=2pt,
    bottom=2pt,
    fontupper=\small\itshape,
    borderline west={1pt}{0pt}{cyan!60!gray},  
    interior style={
        top color=cyan!12!white,               
        bottom color=cyan!25!white             
    },
    drop fuzzy shadow=gray!10                  
]
\small\textit{
\textbf{System Role:} Proactive controller maintaining SLA (response time $<$1s) via adaptive dimmer and server scaling.\\[3pt]
\textbf{Goals:} Keep dimmer near~1.0, and minimize servers (1–3).\\[3pt]
\textbf{Logic:} Analyze trends (use KB tool)~$\rightarrow$~simulate(use WM tool) ~$\rightarrow$~act (adjust dimmer first, scale servers only if persistent overload).\\[3pt]
\textbf{Constraints:} Stable actions, valid JSON output, respect cooldowns, and never violate SLA.
}
\end{tcolorbox}

\paragraph*{Deliberation via Tools.} The Reasoner leverages its tools for evidence-based decision-making. The KB provides a unified access layer to structured and unstructured knowledge such as operational logs, adaptation traces, and historical actions. Any data management or retrieval mechanism may be used, provided it enables queries that return contextually relevant system episodes. The \textbf{World Model (WM)} is used for "what-if" simulations through learned performance models, simulators, or domain specific models etc. 

The Reasoner's deliberation is guided by two query formalisms that interface with its tools:
\[
    q_{KB} = (d_{req}, \lambda) \rightarrow D_{hist} \quad \text{and} \quad q_{WM} = (\phi, \psi) \rightarrow f_{sys}
\]
A query to the Knowledge Base, $q_{KB}$, is composed of a semantic filter $d_{req}$ and a set of structured key-value constraints $\lambda$, returning $D_{hist}$ as a collection of relevant historical event tuples. In our prototype, this query is executed against an in-memory collection of Python objects. Similarly, a query to the World Model, $q_{WM}$, simulates the effect of a proposed action vector $\phi$ on the current system state $\psi$. Implemented in example as an inference call to a Bayesian model, it returns $f_{sys}$, a probability distribution over the predicted next state.

The Reasoner integrates the historical data ($D_{hist}$) and the predictive model ($f_{sys}$) to deliberate over potential actions.

In our example, once condition is stabilized, the Kernel puts the Reasoner back in control whose goal is to restore high fidelity (dimmer=1.0). Reasoner then queries the KB to analyze the recent crisis (the 900\textit{ms} spike) and confirms it was due to high server load and goes to formulates the hypothesis that adding a server might create enough capacity to restore the dimmer. To test this, it queries the World Model with the candidate action $a^*=\mathtt{add\_server}$. The WM simulates this action against the current system state, and its prediction, $f_{sys}$, indicates that response time would remain safely below the 750ms threshold even with the dimmer restored to 1.0.

\paragraph*{Decision and Verification.} The Reasoner formulates a candidate adaptation directive $a^*$. This is sent to the \textbf{Verification Agent (V)}, which acts as a safety backstop, validating the plan against all invariants/constraints $\mathcal{I}$:
\[
V(a^*, \mathcal{I}) \rightarrow \{\text{accept}, \text{reject}\}
\]
For instance, V can be realized as: (a) a lightweight checker for static invariants (e.g., $\texttt{max\_servers} \leq 3$ ), (b) a runtime verification monitor evaluating plans against temporal logic specifications (e.g., LTL rules to prevent undesirable oscillations), or (c) an interface to a formal model checker validating the plan against a formal system model\cite{KNP25, LEUCKER2009293}. In our current evaluation, we employ approach (a). 

As shown in Figure ~\ref{methodology}, reasoning and stabilization paths converge here, and only upon acceptance does the directive proceed to the $EA$ for execution.

\subsection{Meta-Learning}
The outermost loop in Figure ~\ref{methodology} enables the system to learn from its own adaptive behavior to become more effective over time.
\paragraph*{Record (Experience Capture).} After every adaptation cycle, an experience tuple $\epsilon$ is systematically recorded and stored in the $KB$:
\[
\epsilon = (\epsilon_{ctx}, \epsilon_{dec}, \epsilon_{out})
\]
\noindent In our running example, this process captures the full sequence of events. First, the reactive cycle is recorded ($\epsilon_1$): capturing the critical state ($\epsilon_{ctx}$), Fast Controller's \texttt{reduce\_dimmer} action ($\epsilon_{dec}$), and the immediate stabilizing outcome ($\epsilon_{out}$). Subsequently, the proactive cycle is recorded ($\epsilon_2$), capturing the follow-up system context, the Reasoner's \texttt{add\_server} decision, and the final optimized outcome.

\paragraph*{Reflect and Evolve.} Periodically, the \textbf{Meta-Learner Agent (M)} analyzes the accumulated experiences $\{\epsilon_i, ...\}$ from the $KB$ to identify statistically significant patterns ($P$) and then derives an improved strategy ($S'$):
\[
f_{reflect}: KB \rightarrow P \quad \text{and} \quad f_{evolve}: P \rightarrow S'
\]
This new strategy $S'$ is disseminated by updating policies in the Reasoner (modifying it's prompt template $P_{conf}$), tuning thresholds in the Kernel, or enhancing parameters in the World Model. This continuous self-refinement embodies the shift to \textit{Self-Adaptation 3.0}. For instance, after the experiences (high utilization $\rightarrow$ $F$ acts $\rightarrow$ $R$ acts), $f_{reflect}$ identifies a statistically significant pattern $P$: "when server utilization approaches 80\%, the Fast Controller is consistently triggered, followed by an action to add a server." Through $f_{evolve}(P) \rightarrow S'$, the Meta-Learner derives an improved strategy $S'$: it lowers the $\texttt{server\_util\_danger}$ parameter in $\mathcal{O} \subset \mathcal{P}_{conf}$ from 0.90 to 0.80. This update allows the Reasoner to anticipate this condition and proactively $\texttt{add\_server}$, preventing the SLA breach and the need for reactive intervention.

\paragraph*{Prompt Configuration.}
Since the Meta-Learner is implemented using an LLM agent, its reflective and evolutionary behavior is governed through a structured \textit{prompt configuration} that formalizes how experiences are interpreted and transformed into improved strategies. This configuration defines the reflection workflow, evolution logic, and update dissemination, ensuring consistent meta-level reasoning across learning cycles. Formally,
\[
\mathcal{M}_{conf} = \langle \mathcal{R}_m, \mathcal{E}_m, \mathcal{U}_m \rangle
\]
where $\mathcal{R}_m$ represents reflection procedures (e.g., extract trends $\rightarrow$ correlate patterns $\rightarrow$ summarize insights), $\mathcal{E}_m$ defines evolution mappings from patterns to refined strategies, and $\mathcal{U}_m$ governs how updates propagate across the system. Through $\mathcal{M}_{conf}$, the Meta-Learner achieves structured and interpretable continual improvement.

\paragraph*{Example Prompt.}
A skeleton of the prompt for Meta-Learner
\begin{tcolorbox}[
    enhanced,
    colback=cyan!10!white,                     
    colframe=cyan!50!gray,                     
    boxrule=0.4pt,
    arc=2pt,
    left=5pt,
    right=5pt,
    top=2pt,
    bottom=2pt,
    fontupper=\small\itshape,
    borderline west={1pt}{0pt}{cyan!60!gray},  
    interior style={
        top color=cyan!12!white,               
        bottom color=cyan!25!white             
    },
    drop fuzzy shadow=gray!10                  
]
You are an expert systems engineer specializing in adaptive optimization.\\[3pt]
Decide whether to \textbf{store new observations}, \textbf{tune thresholds}, or \textbf{refine templates} based on system evidence.\\[3pt]
Modify only when patterns or parameters clearly justify it; otherwise, return an empty update.\\[3pt]
All outputs must be valid JSON with concise justifications.
\end{tcolorbox}

The meta-learner has access to tools to enforce the required changes and query the current reasoner prompt, knowledge base etc. The Meta-Learner accesses its tools via bidirectional queries 
$q_{ML} = (\eta, \sigma) \leftrightarrow \Delta_S$, 
enabling both retrieval and update of knowledge, configurations, prompt template and thresholds across the system.

\section{Experimental details}
\smallskip
\noindent \underline{\textit{Exemplars}}  We evaluate our framework using two diverse self-adaptive exemplar systems: \textit{SWIM}\cite{swim} and \textit{SWITCH}\cite{switch}. \textit{SWIM} is a web application simulator for resource management under traffic uncertainty, emphasizing elasticity and dimmer control. It monitors response time and server utilization to adapt by adding/removing servers (with a 60s provisioning delay) or adjusting the \textit{dimmer value}, the probability of returning optional content (0–1 range). \textit{SWITCH} is a self-adaptive exemplar for ML-enabled system (MLS). It enables self-adaptation by dynamically switching between machine learning models to balance latency and accuracy. In the evaluated object detection scenario, it selects among vision models based on confidence, size, CPU usage, and response time.
This diversity provides a strong test of our framework’s generalizability. 

\smallskip
\noindent \underline{\textit{LLM selection}}  To test POLARIS, for the Reasoner, we selected a diverse set of LLMs to balance reasoning quality, latency, and cost. We used high-performance proprietary models ~\cite{chiang2024chatbotarenaopenplatform} via API (GPT-5 and Gemini 2.5 Pro for reasoning, Gemini 2.0 Flash for speed). We also used a reasoning based, locally-hosted open-source model (GPT OSS 20B served via Ollama\footnote{https://ollama.com/}). We used the Gemini 2.0 Flash model for the Meta-learner given the task's simplicity and easy availability of the model.

\smallskip
\noindent \underline{\textit{Implementation}}  
POLARIS was implemented in Python. For SWIM, Adapters managed pub-sub communication with telemetry subjects via a NATS server, while the kernel ran asynchronous monitoring and control loops. Response time was derived from telemetry streams; if it exceeded 0.75 s, a reactive script either launched an extra server (if available) or reduced the dimmer level. The Knowledge Base, built with native Python structures, stored buffered telemetry, aggregated observations, and recent actions for rapid lookup. The reasoning layer used an instantiated LLM (local or API) with Chain-of-Thought prompting \cite{chainofthought}, linked to the Knowledge Base and a Bayesian World Model for reasoning under uncertainty. Given the agent’s simplicity, no framework such as CrewAI or LangGraph was used. Control actions from the LLM were published back to SWIM through the execution adapter via NATS topics. A secondary Gemini Flash 2.0 LLM performed meta-learning by analyzing adaptation logs and telemetry summaries, updating prompt files , recalibrating thresholds, and refining reasoning parameters for improved adaptation (through file writing tool-calls).

\smallskip
\noindent \underline{\textit{Hardware Configuration}}
SWIM and SWITCH experiments ran on Ubuntu systems with 16GB RAM (4.7 GHz Intel i7) and 24GB RAM (4.8 GHz Intel Ultra 7), respectively. The local GPT OSS 20B model was served via Ollama on an NVIDIA A6000 GPU.

\smallskip
\noindent \underline{\textit{Experimental Candidates}}
 To ensure a comprehensive evaluation, we compared \textit{POLARIS} against a diverse set of baselines across \textit{SWIM} and \textit{SWITCH}. In SWIM, we considered its built-in rule-based approaches, \textit{Reactive1} and \textit{Reactive2}, along with adaptive approaches such as \textit{Thallium}~\cite{thallium}, \textit{PLA-SDP}~\cite{plasdp}, \textit{Cobra}~\cite{cobrapaper}, and \textit{PLA}~\cite{pla}. For SWITCH, we used \textit{AdaMLS}~\cite{adamls} as the primary baseline. Experiments were conducted using the built-in \textit{Clarknet}~\cite{clarknet_trace, clarknet_paper} trace for SWIM and the \textit{General Object Detection} trace for SWITCH on 500 randomly sampled images from the \textit{COCO 2017} dataset~\cite{coco}. The same Clarknet trace was applied across all SWIM-based baselines for fairness, and selected experiments were repeated three times for statistical reliability (as detailed in RQ1). Based on the results, \textit{Gemini Flash 2.0} was chosen as the Reasoner agent for subsequent ablation studies, having the lowest overall performance among the evaluated LLMs. Ablations were designed to assess the contribution of each framework component, including scenarios without Tools (KB and WM). This does not \textbf{restrict the LLM’s access to information}; rather, KB and WM outputs are provided deterministically. While this eases context retrieval for the LLM, it lessens its agentic capabilities in generalizable scenarios. We aim to show that even with fully autonomous tool use, the framework can match or exceed the deterministic setting.

 \smallskip
\noindent \underline{\textit{Notation for POLARIS Variants}}
For clarity, each POLARIS variant is labeled according to which internal component is inactive. The LLM in the notation refers to the one used for Reasoner. For Meta-learner, we use Gemini Flash 2.0. If a component is absent, it is denoted with a \textbf{--} prefix (e.g., \textbf{--M}). For ex: \textit{POLARIS (Gemini Flash, -M)} denotes a variant running on Gemini Flash based Reasoner without the Meta-Learner but with Tools and Fast Controller active.
\smallskip
Our replication package for more details is available: 
\href{https://github.com/sa4s-serc/POLARIS}{here}

\section{Results}

We define two research questions to evaluate the performance of POLARIS in terms of both effectiveness and efficiency:

\subsubsection*{RQ1} \textit{How does the effectiveness and generalizability of POLARIS compare to existing baseline approaches?}

\begin{table*}[]
\footnotesize
\begin{tabular}{|l|p{1.8cm}|p{2cm}|p{2cm}|p{2.1cm}|p{2.1cm}|}
\hline
\multicolumn{1}{|c|}{\textbf{Approach}}                           & \multicolumn{1}{c|}{\textbf{Optional \%}} & \multicolumn{1}{c|}{\textbf{Late \%}} & \multicolumn{1}{c|}{\textbf{Avg Servers}} & \textbf{Total Utility} \\ \hline
\multicolumn{5}{|c|}{\textbf{Baselines}}                                                                                                                                                                                   \\ \hline
\multicolumn{1}{|c|}{Reactive 1}                                  & \multicolumn{1}{c|}{91.07}                & \multicolumn{1}{c|}{28.78}            & \multicolumn{1}{c|}{2.45}                 & 2647.48                \\ \hline
\multicolumn{1}{|c|}{Reactive 2}                                  & \multicolumn{1}{c|}{88.33}                & \multicolumn{1}{c|}{35.05}            & \multicolumn{1}{c|}{2.40}                 & 1739.88                \\ \hline
\multicolumn{1}{|c|}{Thallium}                                    & \multicolumn{1}{c|}{45.56}                & \multicolumn{1}{c|}{0.30}             & \multicolumn{1}{c|}{2.00}                 & 4658.65                \\ \hline
\multicolumn{1}{|c|}{PLA-SDP}                                     & \multicolumn{1}{c|}{15.56}                & \multicolumn{1}{c|}{0.00}             & \multicolumn{1}{c|}{2.72}                 & 4089.09                \\ \hline
\multicolumn{1}{|c|}{Cobra}                                       & \multicolumn{1}{c|}{89.00}                & \multicolumn{1}{c|}{3.30}             & \multicolumn{1}{c|}{3.00}                 & 5378.00                \\ \hline
\multicolumn{1}{|c|}{PLA}                                         & \multicolumn{1}{c|}{69.10}                & \multicolumn{1}{c|}{0.60}             & \multicolumn{1}{c|}{2.80}                 & 5306.00                \\ \hline
\multicolumn{5}{|c|}{\textbf{POLARIS - Full configurations}}                                                                                                                                                               \\ \hline
\multicolumn{1}{|l|}{POLARIS (GPT-OSS, Full Configuration)}       & \multicolumn{1}{c|}{89.14}                & \multicolumn{1}{c|}{5.42}             & \multicolumn{1}{c|}{2.77}                 & \textbf{5405.98}       \\ \hline
\multicolumn{1}{|l|}{POLARIS (GPT-5, Full Configuration)}         & \multicolumn{1}{c|}{\textbf{95.08}}       & \multicolumn{1}{c|}{6.24}             & \multicolumn{1}{c|}{2.82}                 & \textbf{5445.48}       \\ \hline
\multicolumn{1}{|l|}{POLARIS (Gemini Pro, Full Configuration)}    & \multicolumn{1}{c|}{85.34}                & \multicolumn{1}{c|}{4.86}             & \multicolumn{1}{c|}{2.56}                 & 5093.49                \\ \hline
\multicolumn{1}{|l|}{POLARIS (Gemini Flash, Full Configuration)}  & \multicolumn{1}{c|}{94.4}                & \multicolumn{1}{c|}{6.0}            & \multicolumn{1}{c|}{2.7}                 & 5294.00                \\ \hline
\multicolumn{5}{|c|}{\textbf{POLARIS - Ablations}}                                                                                                                                                                         \\ \hline
\multicolumn{1}{|l|}{POLARIS (Gemini Flash, --M)}                & \multicolumn{1}{c|}{89.31}                & \multicolumn{1}{c|}{18.18}            & \multicolumn{1}{c|}{2.40}                 & 3903.67                \\ \hline
\multicolumn{1}{|l|}{POLARIS (Gemini Flash, --Tools)}             & \multicolumn{1}{c|}{90.71}                & \multicolumn{1}{c|}{5.69}             & \multicolumn{1}{c|}{2.78}                 & 5272.51                \\ \hline
\multicolumn{1}{|l|}{POLARIS (Gemini Flash, --Tools, --F)}       & \multicolumn{1}{c|}{94.77}                & \multicolumn{1}{c|}{12.04}            & \multicolumn{1}{c|}{2.47}                 & 4580.87                \\ \hline
\multicolumn{1}{|l|}{POLARIS (Gemini Flash, --Tools, --M)}       & \multicolumn{1}{c|}{97.82}                & \multicolumn{1}{c|}{13.51}            & \multicolumn{1}{c|}{2.53}                 & 4846.28                \\ \hline
\multicolumn{1}{|l|}{POLARIS (Gemini Flash, --Tools, --M, --F)} & \multicolumn{1}{c|}{97.84}                & \multicolumn{1}{c|}{17.95}            & \multicolumn{1}{c|}{2.45}                 & 4130.68                \\ \hline
\end{tabular}
\caption{Performance metrics for all evaluated  on SWIM with Clarknet Trace. Variants of POLARIS differ by the exclusion (–) of the Meta-Learner (M), Tools (KB + WM), and Fast Controller (F).}
\label{table1}
\end{table*}

Table~\ref{table1} presents the results of applying various baselines on the SWIM exemplar. For \textit{Cobra} and \textit{PLA}, we take results from Moreno et al\cite{cobrapla} study, with their best performing configuration on Clarknet.  From the results, it is evident that POLARIS outperforms the baseline approaches in terms of cumulative utility while maintaining lower response time violations. This improvement can be attributed to the learning-driven adaptation enabled by the interaction between the Reasoner and the Meta-learner. By continuously observing suboptimal decisions that lead to response time spikes, POLARIS refines its decision-making over successive adaptation cycles. As shown in Figure~\ref{fig:fig2}, the full configuration exhibits fewer response-time violations due to the meta-learner’s ability to leverage past observations and past action feedback from the Knowledge Base, enabling faster and more informed adaptation. Unlike the baselines, which rely on complex, mathematically defined strategies requiring extensive design-time modeling, POLARIS leverages \textbf{natural language–based reasoning} to generalize effectively across dynamic and complex environments. 
The framework’s architectural controls mitigate the randomness typical of single LLMs, ensuring consistent behavior across runs. Notably, it delivers comparable performance with different model backends (GPT-5, GPT-OSS:20B, Gemini Pro, Flash), indicating that reliability in LLM-driven systems stems more from design than choosing a specific model.

\begin{figure}[h]
    \centering
    \includegraphics[width=1\linewidth]{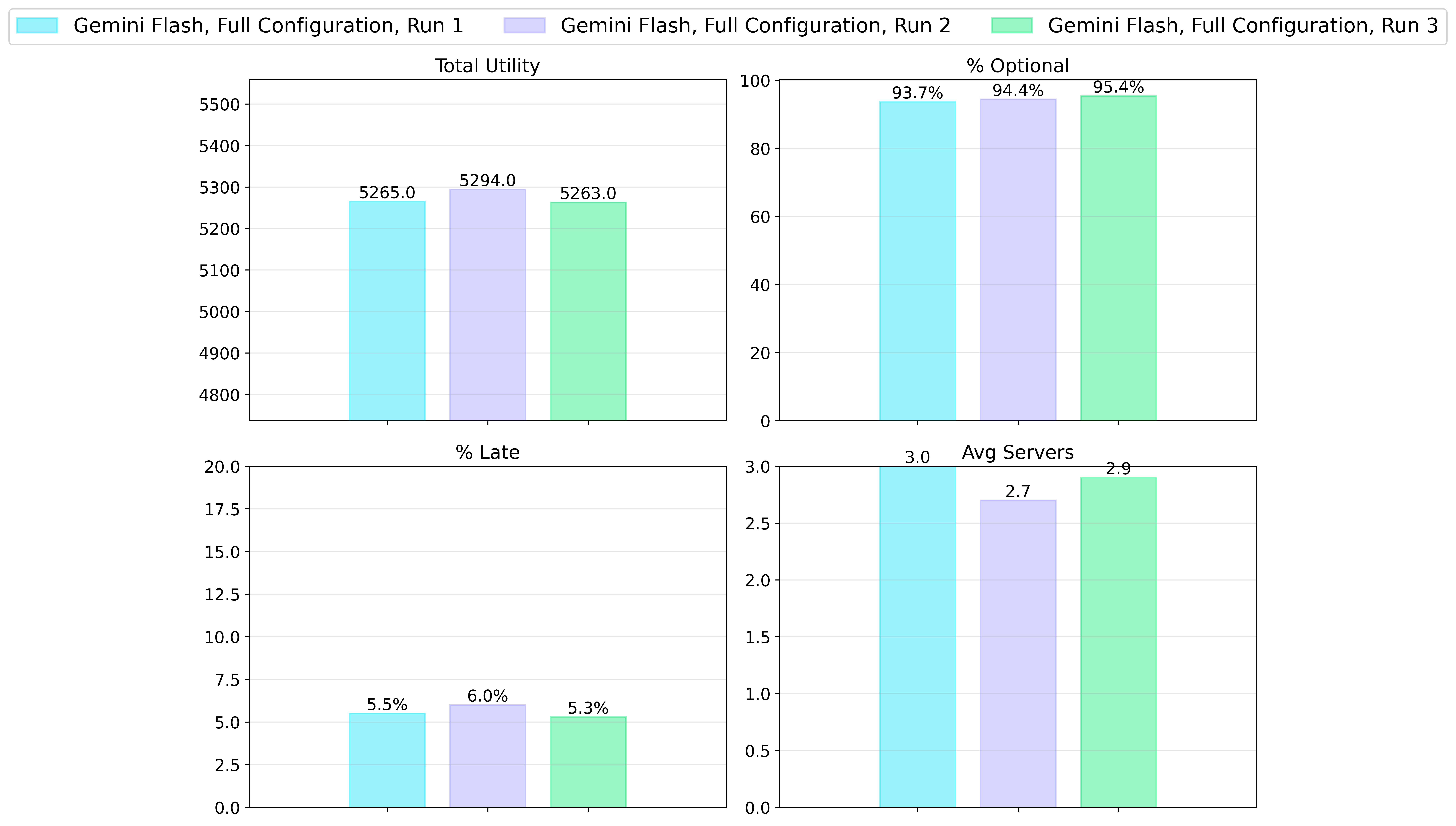}
    
    \caption{Same configuration of POLARIS across 3 runs show minimal variation}
    \label{fig:fig1}
\end{figure}


\begin{figure}[h]
    \centering
    \includegraphics[width=0.8\linewidth]{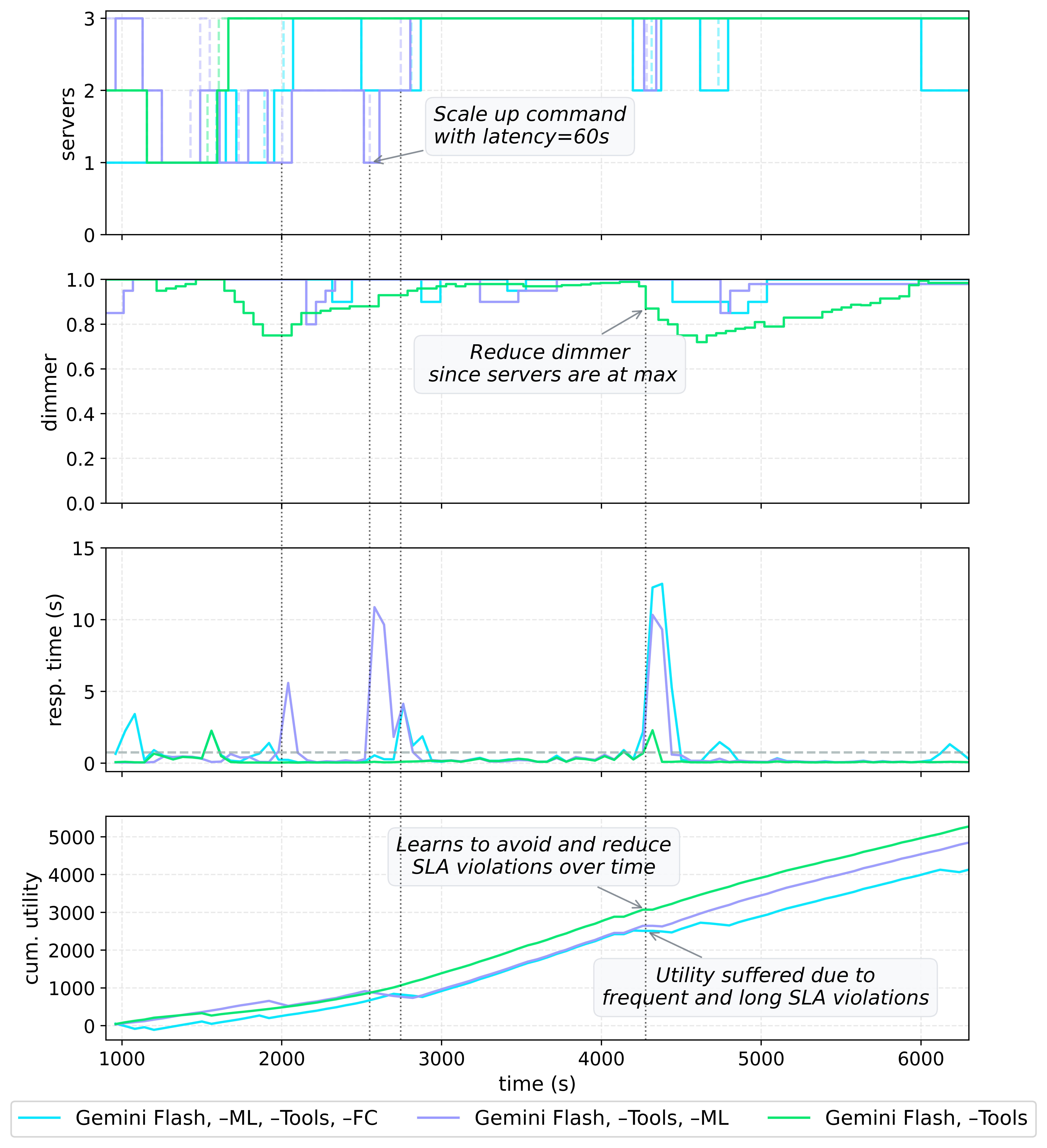}
    \caption{Effect of Fast controller and Meta-Learner in POLARIS}
    \label{fig:fig2}
\end{figure}

\begin{figure}[ht!]
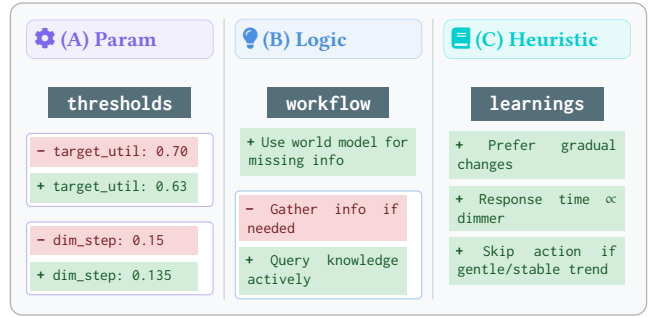

    \centering
    
    \definecolor{accentA}{HTML}{7B68EE}
    \definecolor{accentB}{HTML}{4A90E2}
    \definecolor{accentC}{HTML}{00CED1}
    \definecolor{addbg}{HTML}{D4EDDA}
    \definecolor{delbg}{HTML}{F8D7DA}
    \definecolor{headerbg}{HTML}{546E7A}
    \definecolor{lightgray}{HTML}{F8F9FA}
    \definecolor{addtext}{HTML}{155724}
    \definecolor{deltext}{HTML}{721C24}
    \definecolor{divider}{HTML}{CFD8DC}
    
    \begin{tcolorbox}[
        width=\columnwidth,
        colframe=black!30, boxrule=0.5pt, colback=lightgray, arc=1.5mm,
        left=2mm, right=2mm, top=2mm, bottom=2mm,
        boxsep=0mm,
        shadow={1mm}{-1mm}{0mm}{black!10}
    ]
        \begin{minipage}[t]{0.31\linewidth}
            \centering
            \begin{tcolorbox}[
                enhanced, colback=accentA!10, colframe=accentA!40,
                boxrule=0.3pt, arc=1mm, boxsep=0mm,
                left=1mm, right=1mm, top=1mm, bottom=1mm
            ]
                {\small\textcolor{accentA}{\faCog\ \textbf{(A) Param}}}
            \end{tcolorbox}
            \vspace{1.5mm}
            
            \colorbox{headerbg}{\color{white}\ttfamily\small\bfseries\ thresholds\ }\\[1mm]
            \begin{tcolorbox}[
                enhanced, colback=white, colframe=accentA!50,
                boxrule=0.4pt, arc=0.5mm, boxsep=0mm,
                left=0.5mm, right=0.5mm, top=0.5mm, bottom=0.5mm,
            ]
            \colorbox{delbg}{%
                \parbox{0.85\linewidth}{\color{deltext}\ttfamily\scriptsize\textbf{--}\ target\_util: 0.70}%
            }\\[0.5mm]
            \colorbox{addbg}{%
                \parbox{0.85\linewidth}{\color{addtext}\ttfamily\scriptsize\textbf{+}\ target\_util: 0.63}%
            }
            \end{tcolorbox}
            \begin{tcolorbox}[
                enhanced, colback=white, colframe=accentA!50,
                boxrule=0.4pt, arc=0.5mm, boxsep=0mm,
                left=0.5mm, right=0.5mm, top=0.5mm, bottom=0.5mm,
            ]
            \colorbox{delbg}{%
                \parbox{0.85\linewidth}{\color{deltext}\ttfamily\scriptsize\textbf{--}\ dim\_step: 0.15}%
            }\\[0.5mm]
            \colorbox{addbg}{%
                \parbox{0.85\linewidth}{\color{addtext}\ttfamily\scriptsize\textbf{+}\ dim\_step: 0.135}%
            }
            \end{tcolorbox}
        \end{minipage}%
        \hfill%
        {\color{divider}\vrule width 0.5pt}%
        \hfill%
        \begin{minipage}[t]{0.31\linewidth}
            \centering
            \begin{tcolorbox}[
                enhanced, colback=accentB!10, colframe=accentB!40,
                boxrule=0.3pt, arc=1mm, boxsep=0mm,
                left=1mm, right=1mm, top=1mm, bottom=1mm
            ]
                {\small\textcolor{accentB}{\faLightbulb\ \textbf{(B) Logic}}}
            \end{tcolorbox}
            \vspace{1.5mm}
            
            \colorbox{headerbg}{\color{white}\ttfamily\small\bfseries\ workflow\ }\\[1mm]
            \colorbox{addbg}{%
                \parbox{0.85\linewidth}{\color{addtext}\ttfamily\scriptsize\textbf{+}\ Use world model for missing info}%
            }\\[0.5mm]
            \begin{tcolorbox}[
                enhanced, colback=white, colframe=accentB!50,
                boxrule=0.4pt, arc=0.5mm, boxsep=0mm,
                left=0.5mm, right=0.5mm, top=0.5mm, bottom=0.5mm,
            ]
            \colorbox{delbg}{%
                \parbox{0.85\linewidth}{\color{deltext}\ttfamily\scriptsize\textbf{--}\ Gather info if needed}%
            }\\[0.5mm]
            \colorbox{addbg}{%
                \parbox{0.85\linewidth}{\color{addtext}\ttfamily\scriptsize\textbf{+}\ Query knowledge actively}%
            }
            \end{tcolorbox}
        \end{minipage}%
        \hfill%
        {\color{divider}\vrule width 0.5pt}%
        \hfill%
        \begin{minipage}[t]{0.31\linewidth}
            \centering
            \begin{tcolorbox}[
                enhanced, colback=accentC!10, colframe=accentC!40,
                boxrule=0.3pt, arc=1mm, boxsep=0mm,
                left=1mm, right=1mm, top=1mm, bottom=1mm
            ]
                {\small\textcolor{accentC}{\faBook\ \textbf{(C) Heuristic}}}
            \end{tcolorbox}
            \vspace{1.5mm}
            
            \colorbox{headerbg}{\color{white}\ttfamily\small\bfseries\ learnings\ }\\[1mm]
            \colorbox{addbg}{%
                \parbox{0.85\linewidth}{\color{addtext}\ttfamily\scriptsize\textbf{+}\ Prefer gradual changes}%
            }\\[0.5mm]
            \colorbox{addbg}{%
                \parbox{0.85\linewidth}{\color{addtext}\ttfamily\scriptsize\textbf{+}\ Response time $\propto$ dimmer}%
            }\\[0.5mm]
            \colorbox{addbg}{%
                \parbox{0.85\linewidth}{\color{addtext}\ttfamily\scriptsize\textbf{+}\ Skip action if gentle/stable trend}%
            }
        \end{minipage}
    \end{tcolorbox}
    
    \caption{Meta-learner initiated Reasoner prompt evolution: (A) parameter tuning, (B) logic refinement, (C) heuristic synthesis.}
    \label{fig:prompt-evolution}
\end{figure}

To assess the contribution of individual components, we conducted ablation studies. POLARIS variants without Meta-learner, Fast Controller, or tool integration showed markedly inferior performance, with frequent response time violations. Figure~\ref{fig:fig2} illustrates that incorporating the Meta-learner and Fast Controller significantly improves both cumulative utility and response time stability by enabling proactive and reactive adaptation, respectively. As shown in Figure~\ref{fig:prompt-evolution}, the Meta-learner continuously updates thresholds and prompt configurations based on system observations, enhancing proactive reasoning over time. While the version without tools may appear to perform slightly better in deterministic settings (since tool outputs are directly provided), our objective is to enable autonomous tool use, a key aspect of POLARIS that ensures scalability and generalizability across diverse systems.

\setlength{\textfloatsep}{5pt}
{\begin{table}[h]
\centering
\resizebox{1\columnwidth}{!}{%
\begin{tabular}{@{}lcccc@{}}
\toprule
\textbf{Approach} & 
\textbf{Confidence} & 
\textbf{Response Time (s)} & 
\textbf{CPU Usage (\%)} & 
\textbf{Inference Rate / Switches} \\
& Mean & Median & & (inf/min) \\
\midrule
ADAMLS & 0.729 & 0.132 & 56.85 & 212.78/865 \\
POLARIS\textsuperscript{\textdagger} & 0.688 & \textbf{0.096} & \textbf{48.36} & \textbf{244.19/112} \\
\bottomrule
\end{tabular}%
}
\vspace{2mm}
\footnotesize
\textsuperscript{\textdagger}Gemini Flash with full configuration.
\caption{Performance comparison on SWITCH}
\label{tab:switch}
\end{table}}

To evaluate the generalizability of POLARIS beyond a single domain, we further conducted experiments on the SWITCH exemplar. As summarized in Table~\ref{tab:switch}, POLARIS consistently outperformed ADAMLS across the reported metrics, confirming its ability to generalize effectively across distinct system architectures and workloads. These results demonstrate that POLARIS not only sustains performance under varying operational conditions but also maintains robustness in the presence of uncertainty, highlighting the importance of \textbf{language based reasoning for self-adaptation}.



\subsubsection*{RQ2} \textit{How efficient is POLARIS as a framework, and what overheads does it introduce?}

{
\captionsetup{skip=4pt}
\begin{table}[h]
\centering
\footnotesize
\caption{LLM-wise Latency and Token Usage per Reasoning Decision in POLARIS}
\label{tab:cost}
\resizebox{\columnwidth}{!}{%
\begin{tabular}{l r r r}
\toprule
\textbf{LLM Variant} & \textbf{Input Tokens (avg)} & \textbf{Output Tokens (avg)} \\ 
\midrule
Gemini Flash 2.0 & 8222.97 & 259.66 \\ 
Gemini 2.5 Pro & 13346.32 & 490.74 \\ 
GPT-5 & 17719.02 & 409.12 \\ 
\bottomrule
\end{tabular}
}
\end{table}
}

{
\captionsetup{skip=4pt}
\begin{table}[h]
\centering
\footnotesize
\caption{Performance Efficiency of the POLARIS Framework}
\label{tab:polaris-efficiency}

\begin{tabular}{l r}
\toprule
\textbf{Overhead Type} & \textbf{Average Latency (ms)} \\
\midrule
\textbf{End-to-End Reasoning} & \textbf{7569.57} \\
LLM (Gemini) Call & 3048.31 \\
 World Model (WM) Operation & 11.84 \\
 LLM Tool Call Overhead & 5.23 \\
 Knowledge Base (KB) Query & 2.85 \\
\bottomrule
\end{tabular}
\end{table}

}
In Table~\ref{tab:polaris-efficiency}, we define end-to-end reasoning time as the interval from the first tool call initiated after a decision is generated to the first action generated after that decision cycle. This duration is approximately 7.5 seconds, representing a nominal overhead for the proactive adaptation loop. Since the loop operates at fixed intervals, the impact on overall decision-making is minimal. We argue that the performance and generalizability benefits of POLARIS outweigh this marginal overhead, as any LLM-based reasoning approach inherently incurs inference latency. Beyond the LLM calls, the remaining system components contribute negligible time overheads on the order of milliseconds. Based on Table ~\ref{tab:cost}, while the average token usage for the LLMs does imply a notable operational cost, we argue that this is a manageable trade-off. This cost is clearly outweighed by the significant gains in performance and the overall generalizability that \textit{POLARIS }achieves and is a very manageable level effort for building multi-agentic AI native frameworks.

\section{Discussion}

Our evaluation provides early but compelling evidence that an agentic AI, reasoning-driven approach (what we describe as Self-adaptation 3.0) is both viable and more effective than existing paradigms. This discussion synthesizes our key findings, connects them to the challenges we set out to address, and frames open questions that can guide the community’s next steps.

\noindent \textbf{Generalizable Adaptations with Reasoning:} The evaluation of \textit{POLARIS} demonstrates that it significantly outperforms established baseline approaches in diverse testing environments (seen in \textit{RQ1}). Unlike traditional self-adaptive systems that require extensive design-time modeling, \textit{POLARIS} generalizes effectively across diverse operational contexts by harnessing the power of LLMs for understanding metrics through natural language, bridging the gap between adaptation and reasoning, addressing the imminent \textit{CH1}. 

\noindent \textbf{Learning-Driven Evolution:} The learning capability of \textit{POLARIS} is equally central to its success, as shown in Figures 2 and 3. The system not only reacts to conditions but also records, reflects, and avoids repeating ineffective strategies. Ablation studies (described in \textit{RQ1}) highlight that removing the Meta-Learner or Fast Controller leads to markedly inferior performance, reinforcing the importance of continual reflection and learning. This also contributes towards advancing the vision of lifelong self-adaptation~\cite{gheibiLifelong}. Hence, \textit{POLARIS} takes a tangible step toward realizing \textit{CH2}, demonstrating that adaptation can itself become a learning process.

\noindent \textbf{Coordinated Multi-Agent Intelligence:} Beyond individual learning, \textit{POLARIS} provides a structured blueprint for managing complex adaptation tasks. By balancing short-term goals with long-term planning, it coordinates the Fast Controller, Reasoner, and Meta-Learner in complementary roles. For instance, as depicted in Figure ~\ref{fig:fig2}, the Fast Controller handled rapid response-time spikes, while the Meta-Learner refined thresholds and prompt settings based on prior SLA violations. Removing either one caused a threefold rise in late responses and a drop of over 1,000 in cumulative utility. A similar observation was found for SWITCH (Table ~\ref{tab:switch}). This balance between short-term and long-term reasoning is a leap towards addressing the long-recognized challenge of distributed coordination in self-managed systems~\cite{kramer2007self}, providing early evidence that multi-agent reasoning can achieve both stability and foresight (\textit{CH3}). 

\noindent Applying \textit{POLARIS} to manage a system requires only defining adapters for monitoring and execution, along with specifying prompt templates and the Fast Controller policy, making it flexible and easy to integrate while maintaining respectful efficiency, as observed in \textit{RQ2}. Moreover, the LLM-based reasoning layer supports human-in-the-loop mechanisms through explainable decisions. The Reasoner can also be extended to incorporate richer feedback sources, such as document retrievals or advanced mathematical tools, broadening its capability to derive actionable insights.

\noindent \textbf{A Call to the Community: } We believe POLARIS is a starting point. It opens a research agenda for agentic self-adaptation, prompting key questions: How can we design cognitive feedback loops that complement MAPE-K to enable reasoning about adaptation quality? What new forms of assurance are needed for probabilistic, language-based reasoning? What architectural patterns best support explainable agentic interactions and effective human-in-the-loop collaboration? And finally, how can we evaluate these systems beyond QoS metrics to assess the quality of their learning and reasoning over time?

\section{Threats to validity}
 Our evaluation is subject to some potential threats. \textit{Internal validity} concerns arise from the use of simulated environments; while these are well-established self-adaptive exemplars, they may not fully capture the complexities of real-world systems. Moreover, since the framework relies on LLM-based reasoning, some degree of non-determinism is inherent. Our experimental results in \ref{fig:fig1} show remarkably consistent performance across multiple runs. This suggests that the architectural guardrails, structured prompting, and reflective learning within POLARIS effectively mitigate this risk, leading to reliable and deterministic system-level behavior and further efforts can be put in this direction. \textit{Construct validity} threats stem from the sensitivity of decisions to prompt design, which depends on the developer constructing the managing system. Additionally, tool usage must be explicitly defined, and integrating new tools requires the model to properly interpret tool-calling conventions. Finally, \textit{external validity} is affected by implementation constraints: the Knowledge Base is currently maintained in-memory, and the World Model outputs are implementation-specific. Persisting the Knowledge Base and refining the modeling approach could further strengthen generalization and reasoning robustness.

\section{Conclusion}

We introduced POLARIS, a framework that empowers self-adaptive systems to evaluate and refine their own adaptation strategies over time. By tackling the uncertainties and shifting demands of contemporary software ecosystems, POLARIS sets the stage for the next generation of adaptive software. Its multi-layered design brings together reasoning, learning, and distributed intelligence to support coordinated, adaptive decision-making. Experiments on standard benchmarks demonstrate that POLARIS delivers strong performance and improved generalization. As self-adaptive systems evolve toward more AI-driven paradigms, we expect the fusion of learning and reasoning to become fundamental to their design, analysis, and reliability. POLARIS marks an early step toward this vision, motivating further exploration into systems that not only respond to change but also understand and continually enhance the way they adapt.


\bibliographystyle{ACM-Reference-Format}
\bibliography{references}

\end{document}